\documentclass[journal, twoside]{IEEEtran}
\usepackage{fancyhdr} 
\usepackage{lipsum}
\usepackage[utf8]{inputenc}
\usepackage[T1]{fontenc}

\usepackage{graphicx} 
\usepackage{subfig}
\usepackage{nccmath}
\usepackage{color}
\usepackage{cite}
\usepackage{acro}
\usepackage{scalerel}
\usepackage{tikz}
\usetikzlibrary{svg.path}

\definecolor{orcidlogocol}{HTML}{A6CE39}
\tikzset{
  orcidlogo/.pic={
    \fill[orcidlogocol] svg{M256,128c0,70.7-57.3,128-128,128C57.3,256,0,198.7,0,128C0,57.3,57.3,0,128,0C198.7,0,256,57.3,256,128z};
    \fill[white] svg{M86.3,186.2H70.9V79.1h15.4v48.4V186.2z}
                 svg{M108.9,79.1h41.6c39.6,0,57,28.3,57,53.6c0,27.5-21.5,53.6-56.8,53.6h-41.8V79.1z M124.3,172.4h24.5c34.9,0,42.9-26.5,42.9-39.7c0-21.5-13.7-39.7-43.7-39.7h-23.7V172.4z}
                 svg{M88.7,56.8c0,5.5-4.5,10.1-10.1,10.1c-5.6,0-10.1-4.6-10.1-10.1c0-5.6,4.5-10.1,10.1-10.1C84.2,46.7,88.7,51.3,88.7,56.8z};
  }
}

\newcommand\orcidicon[1]{\href{https://orcid.org/#1}{\mbox{\scalerel*{
\begin{tikzpicture}[yscale=-1,transform shape]
\pic{orcidlogo};
\end{tikzpicture}
}{|}}}}

\ifCLASSINFOpdf
\else
\fi
\DeclareAcronym{acm}{
  short = ACM ,
  long  = Association for Computing Machinery ,
  sort  = ACM,
}

\DeclareAcronym{IT}{
  short = IT ,
  long  = Information Technologies ,
  sort  = IT,
}

\DeclareAcronym{TF-IDF}{
  short = TF-IDF,
  long  = {Term Frequency - Inverse Document Frequency},
  sort  = TF-IDF,
}

\DeclareAcronym{BERT}{
  short = BERT,
  long  = {Bidirectional Encoder Representations from Transformers},
  sort  = BERT,
}

\DeclareAcronym{LDA}{
  short = LDA,
  long  = {Latent Dirichlet Allocation},
  sort  = LDA,
}

\DeclareAcronym{RDS}{
  short = RDS,
  long  = {Amazon Relational Databases},
  sort  = RDS,
}

\DeclareAcronym{EC2}{
  short = EC2,
  long  = {Amazon Elastic Compute Cloud},
  sort  = EC2,
}

\DeclareAcronym{EBS}{
  short = EBS,
  long  = {Amazon Elastic Block Store},
  sort  = EBS,
}

\usepackage{hyperref} 
\begin{document}

 \setcounter{page}{1}
%
\title{COMPARISON OF INFORMATION RETRIEVAL TECHNIQUES APPLIED TO IT SUPPORT TICKETS}
%
%
%

\author{Leonardo Santiago Benitez Pereira \orcidicon{0000-0001-9429-7308}\,, \IEEEmembership{Student Member, IEEE}, Robinson~Pizzio\orcidicon{0000-0002-0034-4888}\,, \IEEEmembership{Member, IEEE}
        and~Samir~Bonho
\thanks{Leonardo Santiago Benitez Pereira is with Skaylink, Lithuania. e-mail:lsbenitezpereira@gmail.com}
\thanks{Robinson Pizzio and Samir Bonho are with the Electronics Department at Federal Institute of Santa Catarina, Brazil. e-mail:robinson.pizzio@ifsc.edu.br samir.bonho@ifsc.edu.br}
}

%

\maketitle

\begin{abstract}

Institutions dependent on IT services and resources acknowledge the crucial significance of an IT help desk system, that act as a centralized hub connecting IT staff and users for service requests. Employing various Machine Learning models, these IT help desk systems allow access to corrective actions used in the past, but each model has different performance when applied to different datasets. 
This work compares eleven Information Retrieval techniques in a dataset of IT support tickets, with the goal of implementing a software that facilitates the work of Information Technology support analysts. The best results were obtained with the Sentence-BERT technique, in its multi-language variation distilluse-base-multilingual-cased-v1, where 78.7\% of the recommendations made by the model were considered relevant. TF-IDF (69.0\%), Word2vec(68.7\%) and LDA (66.3\%) techniques also had consistent results. Furthermore, the used datasets and essential parts of coding have been published and made open source. It also demonstrated the practicality of a support ticket recovery system by implementing a minimal viable prototype, and described in detail the implementation of the system. Finally, this work proposed a novel metric for comparing the techniques, whose aim is to closely reflect the perception of the IT analysts about the retrieval quality.
\end{abstract}

\begin{IEEEkeywords}
Information Retrieval, Machine Learning, Natural Language Processing, Support Tickets, Information Technologies. 
\end{IEEEkeywords}

\IEEEpeerreviewmaketitle

\section{Introduction}


\Ac{IT} has already become part of people's daily lives, who are increasingly dependent on it for educational, social, economic, or professional purposes \cite{stair2009}. To some extent, it is expected that technological resources "just work" and are available all the time, so that unavailability or problems with these technological resources end up disrupting the routine.

Within the business/institutional scope, so-called "IT support teams" are responsible for the proper functioning of IT resources \cite{hawari2021}. When a user faces a problem with IT resources, they describe their problem by opening a support ticket in a management system. Then, an IT Analyst is responsible for solving the problem, communicating with the user, and recording in the management system what actions were taken to solve the problem \cite{zhou2016}.

The knowledge accumulated in these databases is a valuable asset for companies, as it can be used to improve the use of digital technologies within the company. However, searching for this information in the management system databases is technically complicated and time-consuming \cite{fialho2006}. Furthermore, in \cite{silva2020}, the authors point out that IT support teams also face problems such as ticket overload, team turnover, use of inadequate and/or outdated technological tools, lack of qualified labor, among others, further complicating the use of these databases.

Many tickets have identical resolutions, so the analyst only needs to identify if a similar problem has already been solved in the past \cite{muni2017}. For this, the analyst can consult colleagues, read conversation histories, search directly in the database of already resolved tickets, among others. This process can consume a lot of the analyst's time, delaying problem resolution and harming the user.

An area of knowledge that can facilitate the use of these databases is Information Retrieval (IR), whose objective is to find documents of an unstructured nature (usually text) that meet an information need, from a large collection of materials ,\cite{manning2008}. 
Such techniques allow searching for support tickets in the database, making it easier for the IT analyst to find the necessary information to solve a new ticket, which \cite{muni2017} argues can save the analyst a lot of effort and, thus, considerably improve the service provided by IT teams.

The project was carried out at the company Skaylink, using a proprietary database where support tickets are described, and the solutions applied to these tickets are indicated. The eleven chosen IR techniques were applied to Skaylink's database, aiming to define the best IR technique for a scenario where, given a new ticket, the system identifies, among the possibilities available in the database, the similar tickets previously solved to facilitate the work of the IT analyst. It was identified that the {Sentence-BERT} technique, in its multi-language variation {distiluse-base-multilingual-cased-v1}, obtained the best performance among the eleven with 78.7\% relevant recommendations. In this context, the contributions of this research are:
\begin{enumerate}
	\item Compare eleven information retrieval techniques specifically in the context of support tickets, a larger number than any other recent work in the same area;
	\item make the dataset used and the code developed available for free;
	\item propose a new metric to compare information retrieval techniques,
	which closely reflects IT analysts' perception of retrieval quality.
\end{enumerate}

This article is organized as follows: Section II presents a literature review in which the natural language processing (NLP) models used and related works are discussed. Details of the methods employed, involving the dataset, the IR algorithms, and the identification of the best technique are presented in Section III. Results and final considerations are presented in Sections IV and V, respectively.

\section{Literature Review}

For a better understanding of the evaluated methods, a brief description of the NLP models used in this work will be given. Then, various recent works using IR techniques in the domain of IT support tickets are compared, which use a wide variety of methodologies and techniques.

\subsection{Techniques Used}

The \Ac{LDA} model is based on a probabilistic approach that assumes each document in a set of documents is a mixture of a fixed number of topics, and each topic is a mixture of words \cite{blei2003}.

In the \Ac{TF-IDF} model, the frequency of a term in the document is weighted by its occurrence in other texts in the set. The value of this weight is high when the term occurs frequently in the document in question and infrequently in the other documents in the set \cite{tf-idf}. Similarly, the {Best Match 25} (BM25) technique also considers the frequency of keywords in the document but penalizes very long documents \cite{Robertson2009}.

The {Word2Vec} model \cite{word2vec} is a representation based on artificial neural networks (ANN) that relies on the premise that similar words have similar contexts, known as distributional similarity. One of its extensions, {Doc2Vec}, allows representing not only isolated words but entire documents in vector form \cite{doc2vec}.

The \Ac{BERT} \cite{bert} model is pre-trained on an extensive dataset and was pioneering in introducing the concept of bidirectional training, i.e., it takes into account the entire set of words in a sentence simultaneously.

The {Sentence-BERT} is an extension of the BERT model, which focuses on producing vector representations for sentences or expressions, as opposed to representing words or individual units \cite{sentencebert}. The idea of this model is to capture the semantics and context of sentences, allowing the evaluation of semantic similarity between them. For this, a siamese neural network with a pooling operation at the output is used. Additionally, the model is trained on a dataset composed of pairs of sentences annotated with the relationship between them, so that it is "forced to learn" how to properly represent the two sentences. Both \Ac{BERT} and {Sentence-BERT} can handle new words (outside the training dictionary), unlike the other techniques compared in this work.

\subsection{Related Work}

The work of \cite{muni2017} used pre-processed text from the ticket description with techniques such as lemmatization and stop word removal, then applied \Ac{TF-IDF} and dimensionality reduction techniques to obtain a vector representation of the ticket, and finally compared the vectors using Cosine Similarity. The data annotation for training was performed with an existing system. 
The evaluation of the models was done by comparison with the existing system, using both the textual comparison method {SS-Evaluator} and a manual analysis by a support analyst.

In \cite{feng2022}, embeddings from the \Ac{BERT} network and its derivatives (specifically: RoBERTa, DistilBERT, and DistilRoBERTa) were used to enable semantic search of tickets. Additionally, supervised models were used to classify the group/department of the company that should be responsible for the ticket and also the analyst who should handle the ticket. For all tasks, the top-k accuracy metric was used for evaluation.

The High Performance Computing Systems Research Center of the {Los Alamos National Laboratory} published in \cite{delucia2020} their methods for automatically classifying tickets and suggesting similar tickets. They used 70,000 tickets, whose text was pre-processed (stop word removal, conversion to lowercase, among others) and then vectorized (using 3 different techniques: {Latent Dirichlet Allocation}, {Latent Semantic Analysis}, and {Doc2Vec}). The system was initially evaluated by comparison with two existing systems (the "more like this" functionality of the {ElasticSearch} software and an expert system that compares the percentage of common words); thus, 200 tickets were used for a qualitative manual evaluation by an expert, and no quantitative evaluation was performed.

When there is no prior system for comparison (e.g., \cite{muni2017} and \cite{delucia2020}), it is common to use a small evaluation set. In \cite{sund2021}, only 5 tickets were used, but the recommendations were evaluated in two different dimensions: whether the recommendations belonged to a similar area/category (examples of categories in this work are "video editing" and "graphical interface") and whether the recommendations had the same functional characteristic (in this work, functions such as "start", "save", among others were used). Based on this evaluation methodology, 6 vectorization techniques were compared (two variations of {Word2Vec}, two variations of {Doc2Vec}, \Ac{TF-IDF}, and \Ac{BERT}, using only non-retrained models), using the metrics of {total score} and {average score} in each of the two dimensions, where {Doc2Vec} and \Ac{BERT} obtained the best results.

Beyond the retrieval of similar tickets, there are various applications of Machine Learning and related areas to facilitate the resolution of support tickets. In the work of \cite{hawari2021}, for example, 1585 tickets were used to train a model that classified the ticket into one of 13 categories. The concatenation of the title, description, and comments fields was used as input; additionally, pre-processing with stemming and \Ac{TF-IDF} was used to represent the tickets, and 4 models were tested: a rule-based system, J48, {Naive Bayes}, and Sequential Minimal Optimization (SMO). In \cite{capivaraCymaen2022}, also conducted at the company Skaylink, a model is presented that classifies the ticket into 7 categories (using a fully connected 6-layer ANN), but no information retrieval techniques were applied. Beyond simple classification, the work of \cite{zangari2023} performs multi-level classification, while \cite{benge2023} performs multi-label classification, assigning one or more categories out of 10 possibilities; both works use a Bert neural network.

Several companies have also developed solutions for their internal ticket management systems, such as Uber, which in \cite{cota2018} presents a system that classifies tickets to define response templates that the analyst can use. The resulting models were evaluated with real users and reduced the ticket resolution time by 10\%.

\section{Methods}

In this section, we present the methodological details used in the development of this article. Details about the dataset used, the IR algorithms compared, the definition of metrics, among others, are described in the following subsections. Details about the implementation of each technique, as well as the identification of the best among the selected ones, are also described in this section.

\subsection{Dataset}
The dataset used consists of information from 20,356 support tickets submitted between the years 2017 and 2022, recorded during the provision of services for a Skaylink client company. The data was anonymized before the realization of this work, so no personal information was present.

Each ticket is described by nine variables: {external\_ID} (ticket identification in the management system), {title} (ticket title, as informed by the user who opened the incident), {description} (ticket description, also informed by the user), {category} (incident category), {date\_open} (ticket opening date), {date\_close} (ticket closing date), {location} (office the user belongs to), {solution} (solution applied to the ticket), and {analysts} (group of analysts responsible for resolving the ticket). Upon receiving the ticket, the analyst does not receive the variables {category}, {date\_close}, and {solution}, which are added to the records only after the ticket is closed.

For the purposes of this work, the {title} and {description} fields were concatenated and all other fields were discarded, so the system uses only the information provided by the user at the time of incident opening. The top 15 categories present in the dataset, ordered by the number of incidents, are as follows: {Fileservices}, {Active Directory}, {Computer Services}, {Access Control}, {End-Of-Life}, {O365}, {Exchange}, {Create Account}, {Data Center}, {Identity Management}, {Fileshared}, {Telecom}, {Printer}, {Software general}, and {Security}.

The tickets are written primarily in English (207 out of the 300 selected tickets); however, many are written in Portuguese (51 tickets), German (22 tickets), Spanish (19 tickets), French (1 ticket), and potentially the system in operation will receive tickets in other languages. It is noticeable that the dataset is highly unbalanced, and many tickets contain grammatical errors and abbreviations. The concatenation of {title} and {description} totals, on average, 224 characters. For illustration purposes, Table \ref{quadro:example} presents a typical ticket (the {external\_ID} field, names, and dates have been used fictitiously to preserve the identity of the original user).

\begin{table}[!ht]
	\centering
	\caption{Example of a typical ticket.}
	\begin{tabular}{ll}
		\hline
		\hline
		Column & Value \\ \hline
		{{external\_ID}} & {ABC123456} \\
		{{title}} & {File Access} \\
		{{description}} & \begin{tabular}[t]{@{}c@{}} Good morning. I need access to Leonardo Benitez's \\ computer. He has been dismissed from the company and \\ I need the files left on his desktop. These are control \\ spreadsheets and also emails. He has signed the \\ authorization letter. Thank you \end{tabular} \\
		{{category}} & {Fileservice} \\
		{{date\_open}} & {2022-01-01 10:23:19.000} \\
		{{date\_close}} & {2022-01-02 09:15:56.000} \\
		{{location}} & {BRLM} \\
		{{solution}} & \begin{tabular}[t]{@{}c@{}} The client gained access and copied the files to\\ their machine\end{tabular} \\
		{{analysts}} & {Leonardo Pereira} \\ 
		\hline
		\hline
	\end{tabular}
	\label{quadro:example}
\end{table}

The process of identifying similar tickets is also called data labeling \cite{faceli2011}; in this work, 300 tickets were labeled, chosen by the following methodology: from each of the 10 most frequent categories, 30 "representative" tickets were manually chosen, i.e., those that describe a single problem completely.
The 300 tickets were randomly divided into 3 subgroups of 100 tickets, to facilitate labeling. For each ticket, the five most similar tickets (within the subgroup) were manually indicated. 
The decision to indicate exactly 5 similar tickets was made to simplify and make predictable the way the support analyst interacts with the system: many similar tickets are not needed to help solve a new problem, while at least some approximately similar tickets are more useful than none.

\begin{figure}[!ht]
	\centering
	\includegraphics[width=0.70\linewidth]{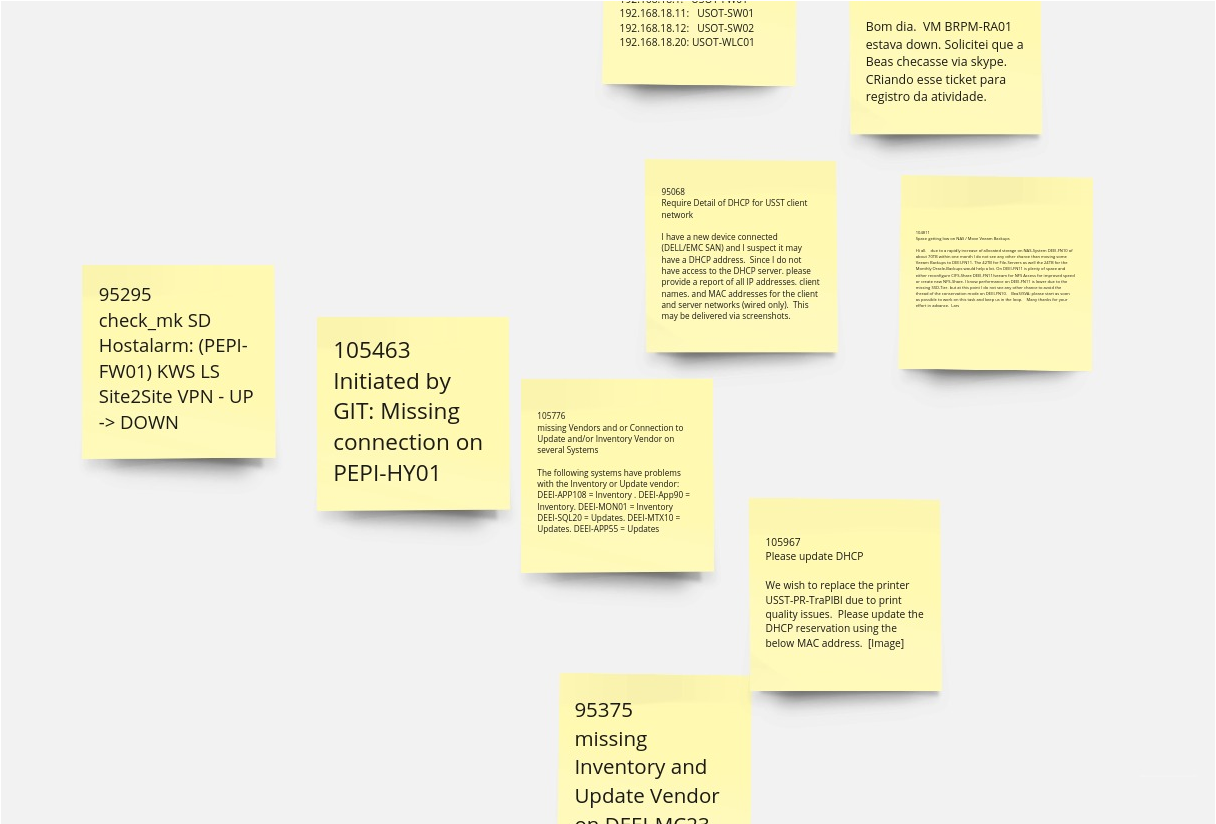}
	\caption{Tool used for labeling.}
	\label{fig:board-miro}
\end{figure}

To simplify labeling, the graphical tool Miro \cite{miro} was used, placing the text of each ticket on a card arranged in a two-dimensional plane (Fig.~\ref{fig:board-miro}), so similar tickets are positioned close to each other. The aforementioned procedure was carried out with 3 analysts labeling the data independently, each responsible for a subgroup, to reduce labeling bias.

The set of 300 tickets used is available at \cite{capivaraRetrieval300Data}. 
Before the data was published, all personal and/or sensitive information was removed (replaced by a tag indicating the original content, for example: the phrase "this text was written by Leonardo" is converted to "this text was written by [NAME]"). The removal was performed in three steps: first, the Machine Learning-based tool {AWS Comprehend PII Removal} was used; then, a sequence of custom regular expressions was applied; finally, all tickets were manually verified.

\subsection{Chosen Techniques}
From the study of IR techniques available to meet the objectives, it was chosen to compare several techniques that are representative of the main existing approaches to the problem:

\begin{itemize}
	\item representing traditional approaches that are highly dependent on the developer's domain knowledge, an Expert System was developed;
	\item representing well-established statistical approaches in the IR field, \Ac{TF-IDF} was used;
	\item representing well-established probabilistic approaches in the IR field, BM25 and \Ac{LDA} were used;
	\item representing approaches of neural networks with non-contextual embeddings, {Word2vec} trained on an English language dataset, {Word2vec} entirely trained with the Skaylink database, and its contextual variation {Doc2vec} also trained with the Skaylink database were used;
	\item representing approaches of neural networks with contextual embeddings, \Ac{BERT} trained with multi-language data, {Sentence-BERT} trained with multi-language data, {Sentence-BERT} trained with English language data only, and {Sentence-BERT} initially trained with multi-language data and then retrained with the Skaylink database were used.
\end{itemize}

Additionally, the technique of random selection - in which calls are selected randomly - was chosen to facilitate the interpretation of the obtained values.

\subsection{Implementation of Techniques}
All techniques were implemented using the {Python} programming language.
The vector generated by each technique was compared using the Cosine Similarity metric, defined as Eq.~(\ref{eq:cossine}). The only exceptions were the expert system and BM25, which used their own similarity metrics. The system then recommends the calls with the highest similarity among the possible recommendations.

\begin{equation} \label{eq:cossine}
	\text{Cosine Similarity}(A,B) = \frac{\mathbf {A} \cdot \mathbf {B}}{\|\mathbf {A} \|\|\mathbf {B} \|}
\end{equation} 

All techniques were implemented following an object-oriented programming structure, so they exposed the same generic interface. Thus, they were evaluated following the same implementation of metrics and methodologies.

\subsubsection{Expert System}
A system was developed that, based on the presence or absence of certain terms, generated a set of "labels" for each document. 116 IT jargon terms were used as terms, complemented with 141 synonyms. To increase the number of labels identified for each document, the text was preprocessed as follows:

\begin{itemize}
	\item converting to lowercase;
	\item removing special characters $-$,  $.$,  $,$,  $!$,  $?$,  $\text{\_}$, and $*$;
	\item converting characters to their closest representation in unicode.
\end{itemize}

After preprocessing, the set of resulting labels was compared using the Jaccard Similarity metric, presented in Eq.~(\ref{eq:jaccard}).

\begin{equation} \label{eq:jaccard}
	\text{Jaccard Similarity}(A,B) = \frac{|A \cap B|}{|A \cup B|}
\end{equation} 

\subsubsection{\Ac{TF-IDF}}
For the \Ac{TF-IDF} technique, the implementation of the {Sklearn} library (in its version 0.23.2) was used. The dictionary was calculated based on the 20056 calls that were not used for evaluation (i.e., out of the total 20356 calls, 20056 were used to calculate the dictionary and 300 for evaluating the model). The same preprocessing as the expert system was used, along with the removal of stop words for the English language. The top 500 most frequent words from the dictionary were used, resulting in a 500-dimensional vector.

\subsubsection{BM25}
The implementation of the {rank\_bm25} library (version 0.2.2) was adopted. The same preprocessing as the expert system was used, and the default parameters of $k1=1.5$, $b=0.75$, and $epsilon=0.25$ were kept. Since the BM25 technique requires its own comparison metric, the implementation of the library was also used.

\subsubsection{\Ac{LDA}}
To implement the \Ac{LDA} technique, the {Gensim} library (version 4.3.0) was chosen. The same preprocessing as the expert system was used, and the technique was configured for 300 topics. This dimension was chosen because it is the same value as the vectors from {Word2vec}, making the comparison fairer, besides being the value used in the work of \cite{muni2017} and being a usual value in the literature.

\subsubsection{{Word2vec} and derivatives}
The implementations from the {Gensim} library were adopted. For the {Word2vec} technique, a model trained on the {Google News} dataset, which contains approximately 100 billion words from the English language, was used. This model produces a 300-dimensional vector. A new {Word2vec} network was also trained using only the 20056 calls that were not used for evaluation (which totaled approximately 1.6 million words), also with 300 dimensions. The same preprocessing as the expert system was maintained.

{Doc2vec} was trained with the same 20056 calls, techniques for preprocessing data, and output vector dimensions. Training was performed with parameters of $window=10$, $min\_count=1$, and $epochs=100$, obtained after briefly testing various values and measuring the accuracy obtained. A pre-trained model was not used because there was no available model in the {Gensim} library, and another reliable source was not found to obtain a model.

\subsubsection{BERT and derivatives}

For the original BERT model, the {BERT-base-multilingual-cased} implementation from the {HuggingFace} library (version 4.25.1) was adopted, which was trained with 104 languages, taking the [CLS] special token output as the embedding, which has a dimension of 768 elements. The {SentenceTransformers} library was used for {Sentence-BERT} models, version 2.2.2. The multi-language model used was {distiluse-base-multilingual-cased-v1}, which was trained with 15 languages (including English, German, Spanish, and Portuguese), providing a 512-dimensional vector. When retraining the network with the Skaylink data, the same {distiluse-base-multilingual-cased-v1} model was used as the base, and it was trained for 3 epochs with the 20056 calls not used for evaluation. The English language model chosen was {all-mpnet-base-v2}, which provides a 768-dimensional vector.

The BERT model and its variations tend to yield better results when the text is not preprocessed \cite{transformersBook}; therefore, the original text was used.

\subsection{Identification of the Best Technique}

The similarity between the calls was calculated based on the positions of the cards, indicated during the data labeling, which were exported to CSV files (one file for each subgroup). For each model, these CSV files were loaded one by one, and the model under evaluation gave 5 recommendations for each of the 100 calls from each subset. The evaluation was performed using the precision metric, presented in Eq.~(\ref{eq:precision}). To measure precision, the 5 calls manually indicated during labeling were considered as "relevant documents," that is, the 5 calls closest in the two-dimensional plane.

\begin{equation} \label{eq:precision}
	\text{Precision} = \frac{\text{Number of relevant recommended items}}{\text{Total number of recommended items}}
\end{equation} 

After that, the average precision\cite{manning2008} among the 3 labeled sets was calculated, and the best technique was chosen as the one with the highest average precision.
The recall metric\cite{manning2008} was not used because, as it was decided that each technique would retrieve five calls and there were also always five calls considered relevant, recall and precision inherently have the same value.
Ranked metrics such as Recall@k \cite{manning2008} were not used because - for this specific IT support case - they do not adequately represent the utility of the system to the support analyst: since each call is briefly described, the analyst can scan the 5 calls for any useful information within a few seconds, making the order of documents less relevant.

Additionally, the "at least one accuracy" metric was created (formalized in Eq.~(\ref{eq:acc-pelo-menos-um}), where $N$ is the number of calls, $y_i$ is the set of relevant calls, and $\hat y_i$ is the set of recommended calls), with the goal of having a metric that better reflects the analyst's perception of the system quality. This metric considers a "hit" if any of the five retrieved calls is indeed a relevant call. For example, if given a support call about "broken mouse," the system retrieves the previous calls "lost my mouse," "mouse not working," "can't log in," "air conditioner too cold," and "password recovery," such retrieval is considered a hit because at least the second call is relevant.

\begin{equation} \label{eq:acc-pelo-menos-um}
	\text{Accuracy}_{\text{at least one}}(y, \hat y) = \frac{\sum_i^N \lambda(y_i, \hat y_i)}{N}
\end{equation}
where
\begin{equation} \label{eq:acc-pelo-menos-um-2}
	\lambda(a,b) = \big\{^{1, |a\cap b|>0}_{0}
\end{equation}

\subsection{Prototype}

\begin{figure}[ht!]
    \centering
    \includegraphics[width=0.9\linewidth]{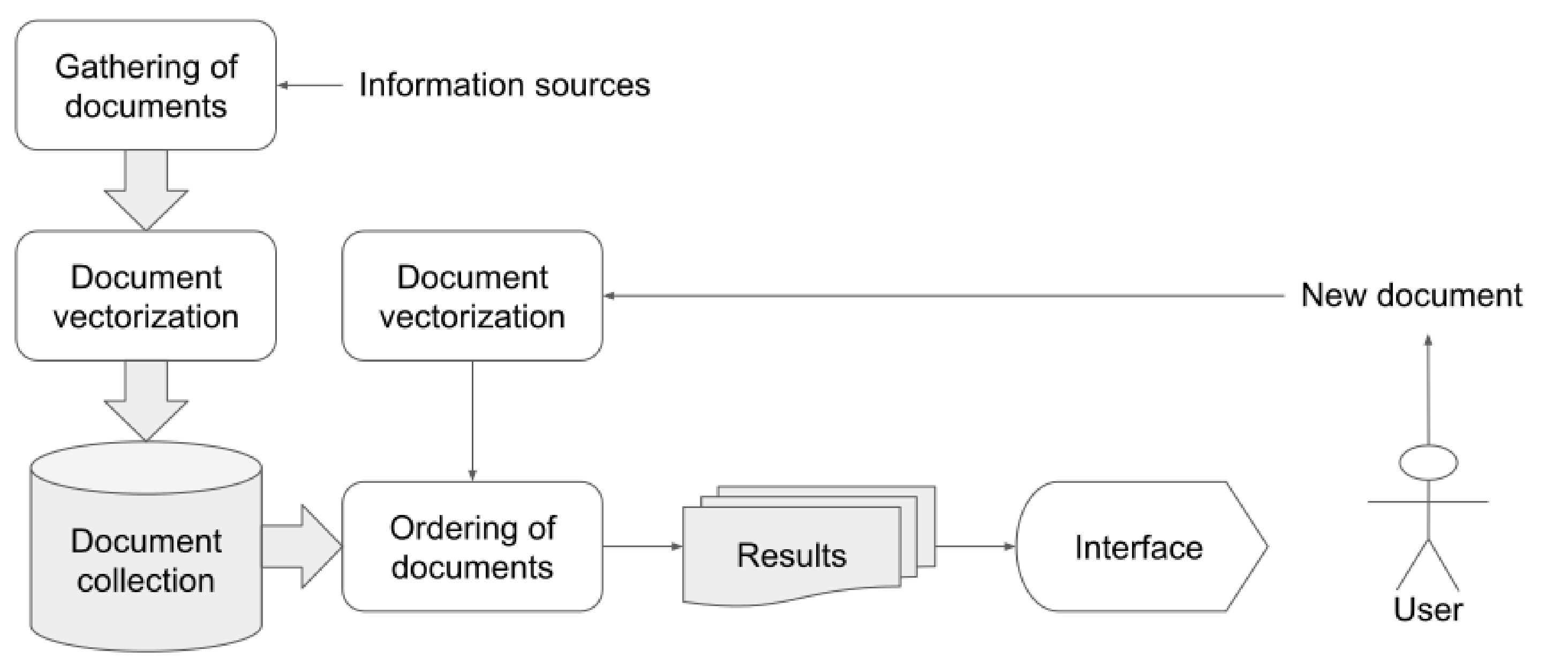}
    \caption{Prototype architecture.}
    \label{fig:arquitetura-implementada}
\end{figure}

Using the technique identified earlier as the most suitable one, a prototype software was implemented. The system architecture followed the proposal from \cite[pg. 907]{OxfordLinguistics} and was implemented as shown in Fig.~\ref{fig:arquitetura-implementada}. This architecture has two distinct flows: document collection and processing of new queries or documents.

During document collection, all 20356 calls from the database are registered, and their respective texts are vectorized. Once vectorized, these documents are stored in a database and thus become available for future queries.

During the processing of new documents, their vector representation is calculated; then, the 100 most recent registered calls are searched, from which the 5 most similar calls are selected and presented to the analyst through a graphical interface. This restriction to the 100 most recent calls is for practical reasons, to avoid system overload, but a future implementation of the system may allow searching the entire database. Additionally, the new call is also stored in the system's database, allowing its use in future recommendations. The prototype provides a button to allow the analyst to give feedback, indicating whether the recommendations were helpful or not, and it was used in Skaylink's daily work.

\section{Results}

The results were summarized in Table \ref{quadro:results}. The time required to perform 100 recommendations (1 subgroup) was also included, calculated by the average of 3 executions (one in each subgroup), including the time to load the model from disk and not including the possible training time when executed on an Acer Aspire notebook. It can be observed that the multi-language Sentence-BERT technique presents the best result, with 35.1\% precision and 78.7\% at least-one accuracy (meaning, three out of four times the system recommends at least one previous call similar to the call being analyzed). Sentence-BERT is also the most recent technique, published in 2019 \cite{sentencebert}, so it was expected to have the best results.

Two techniques surprised by their low precision: multi-language BERT (17.2\%) and Doc2vec (5.8\%). Although both obtain better results than random selection (5.5\%), the literature indicates that the use of these techniques usually results in precisions comparable to classical techniques like TF-IDF (29.6\%). For Doc2vec, a possible explanation is that the training resulted in overfitting to the dataset, given the low number of calls used, 20056, and because it was not based on a previously trained model (i.e., training was done "from scratch"). Regarding multi-language BERT, a possible explanation was the choice of the [CLS] special token as the embedding, as there are other possible methods to extract embeddings from a BERT network.

Regarding the effect obtained by retraining the ANNs (from an existing base model), there were divergent results. Although retraining Word2vec improved its precision, retraining multi-language Sentence-BERT did not achieve the same result, and both had similar performance. This fact can be explained because BERT networks need a large volume of data to be trained, and the use of small datasets can even cause the network to "forget" part of what it learned previously.

\begin{table}[!ht]
	\centering
	\caption{Comparison of implemented techniques.}
	\begin{tabular}{llll}
		\hline
		\hline
		Name & $\text{Accuracy}_\text{alo}$ & $\text{Precision}$ & $\text{Time(ms)}$ \\ \hline
		BM25 & 59.0\% & 23.7\% & 258 \\
		BERT multi-language & 50.0\% & 17.2\% & 12781 \\
		Doc2vec & 27.3\% & 5.8\% & 933 \\
		LDA & 66.3\% & 20.9\% & 833 \\
		Random selection & 26.0\% & 5.5\% & 199 \\
		Sentence-BERT English & 74.3\% & 30.1\% & 10601 \\
		Sentence-BERT multi-language & 78.7\% & 35.1\% & 6411 \\
		Sentence-BERT retrained & 78.7\% & 32.7\% & 6450 \\
		Expert system & 42.7\% & 17.2\% & 1101 \\
		TF-IDF & 69.0\% & 29.7\% & 672 \\
		Word2vec English & 58.3\% & 23.4\% & 49298 \\
		Word2vec retrained & 68.7\% & 26.2\% & 49590 \\
		\hline
		\hline
	\end{tabular}
	\label{quadro:results}
\end{table}

\section{Final Remarks}
Throughout this work, a comparison of eleven Information Retrieval techniques was carried out, applied to a dataset referring to IT support calls. These techniques included various approaches to IR, making it possible to clearly identify the possibilities to implement a system that, given a new support call, is capable of retrieving similar support calls.

The best result was obtained with the {Sentence-BERT} technique, in its multi-language variation {distiluse-base-multilingual-cased-v1}, where 78.7\% of the recommendations made by the model were considered relevant. The two other variations tested from {Sentence-BERT} presented the second and third best results, followed by the \ac{TF-IDF} technique.

Furthermore, this work sought to contribute to the academic community by providing, free and unrestricted, the dataset used and the implementation of the techniques, as well as proposing a new metric for evaluating IR techniques. These results meet the proposed objectives and guide the development of future work in the field.

When interpreting the results obtained, it may seem at first glance that the evaluation metrics indicate poor results. However, it is important to note that the evaluation methodology was defined strictly, and only 5 out of 99 calls were considered relevant (the 100th call being the one being evaluated). Moreover, the nature of the data itself (short texts, poorly explanatory, and with many technical jargon) makes the application of information retrieval techniques challenging. Nevertheless, all implemented techniques showed better results than random selection, indicating that they were able to capture the semantics of support calls.

As a complementary result, it was confirmed that the {Sentence-BERT} network presents better results in IR than the original BERT, as presented in the work of \cite{sentencebert}. This superiority was maintained in all variations of {Sentence-BERT} tested and for all evaluation metrics used.

It is worth noting the positive result of the \ac{TF-IDF} technique, which is simple to implement and computationally fast. The technique also proved to be robust in all experiments conducted, consistently presenting results, while other techniques did not perform equally well under all conditions. Thus, it is possible that the final system will be implemented with \ac{TF-IDF} instead of {Sentence-BERT}.

Finally, it is worth mentioning that the expert system obtained considerably poor results. Despite its simplistic implementation, this demonstrates the difficulty of implementing manually crafted IR systems, justifying the use of more advanced Machine Learning techniques.

Based on the results obtained, it is possible to continue the work with, for example, the following topics:
not restricting the prototype to search among only the last 100 registered calls;
using a database with native support for vector similarity search (such as the ElasticSearch software);
Development of variations of the test dataset, identifying under which conditions each of the techniques presents the best result;
testing other techniques (such as FastText or GPT3);
testing the {Sentence-BERT} model retrained for more than 3 epochs and on a larger dataset;
combining embedding extraction techniques with classification and ranking methods to improve results;
integrating the prototype with the Skaylink company's management system, facilitating the use of the system;

\section*{Acknowledgments}

The authors would like to thank Skaylink company for providing the data used in this study, as well as for their support during the implementation and evaluation process of this work.

\bibliographystyle{ieeetr}  

\ifCLASSOPTIONcaptionsoff
  \newpage
\fi


%

\begin{IEEEbiography}[{\includegraphics[width=1in,height=1.25in,clip,keepaspectratio]{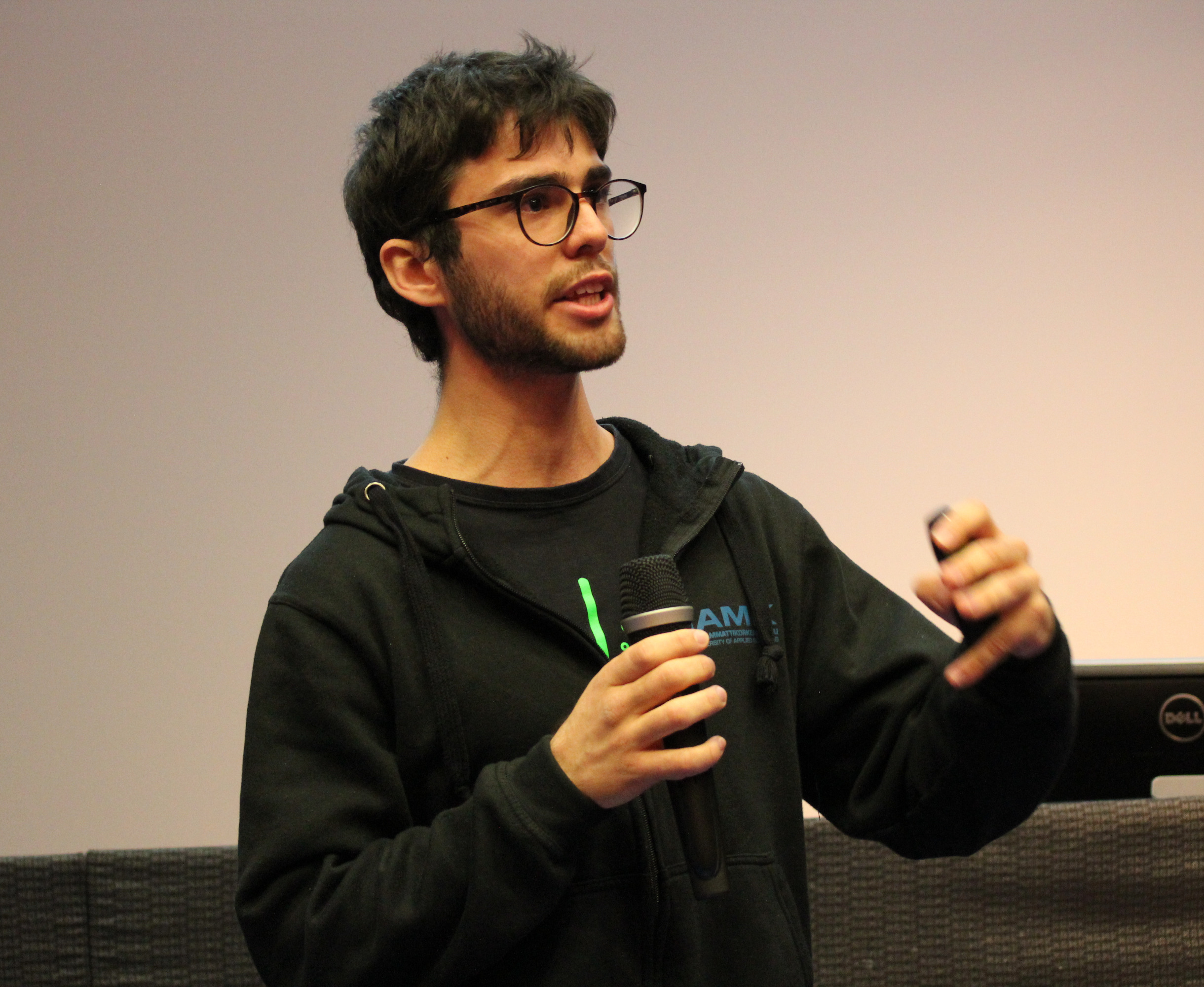}}]{Leonardo Santiago Benitez Pereira} 
received the B.Sc. degree in electronics engineering from the Federal Institute of Santa Catarina (IFSC), in 2022. He has extensive knowledge in software development, has carried out projects in different application areas, always with Machine Learning and data-driven solutions at their core. He currently works at Skaylink as a Machine Learning Engineer.
\end{IEEEbiography}

\begin{IEEEbiography}[{
\includegraphics[width=1in,height=1.25in,clip,keepaspectratio]{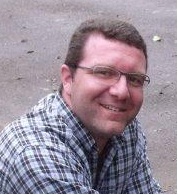}
}]
{Robinson Pizzio} 
received the B.Sc. and M.Sc. degrees in electrical engineering from the Pontifical Catholic University of Rio Grande do Sul (PUCRS), Brazil, in 1995 and 1998, respectively, and the Ph.D. degree from the Federal University of Santa Catarina (UFSC), Brazil, in 2018. He has worked as an assistant professor at PUCRS and University of Caxias do Sul (UCS). Since 2013 he has been an associate professor at IFSC, and since 2022 is the Director of the IFSC's Innovation Hub.
\end{IEEEbiography}


\begin{IEEEbiography}[{\includegraphics[width=1in,height=1.25in,clip,keepaspectratio]{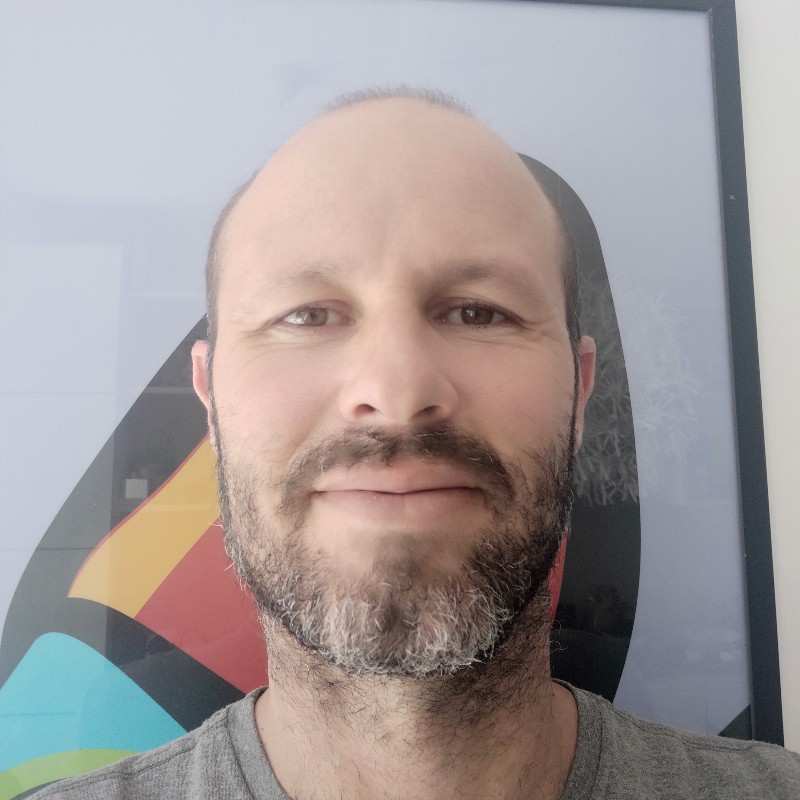}}]{Samir Bonho}
received the B.Sc. and the M.Sc. degree in electrical engineering from the UFSC, in 2004 and 2006, respectively. He has experience in Biomedical Engineering, with emphasis on digital signal processing and data transmission over IP networks. He is currently a professor with the Electronics Department at the IFSC, Florianópolis Campus. He is father of Yannis and Isadora. 
\end{IEEEbiography}
\end{document}